\documentclass[twocolumn,citeautoscript,showpacs,preprintnumbers,amsmath,amssymb,prb,floatfix,footinbib]{revtex4}
\usepackage{graphicx}
\usepackage{dcolumn}
\usepackage{bm}
\usepackage{times}
\begin{document}
\title{Commensurate antiferromagnetic ordering in Ba(Fe$_{1-x}$Co$_{x}$)$_{2}$As$_{2}$ \\determined by x-ray resonant magnetic scattering at the Fe K-edge}
\author{M. G. Kim\textsuperscript{1}, A. Kreyssig\textsuperscript{1}, Y. B. Lee\textsuperscript{1}, J. W. Kim\textsuperscript{2}, D. K. Pratt\textsuperscript{1}, A. Thaler\textsuperscript{1}, \\S. L. Bud'ko\textsuperscript{1}, P. C. Canfield\textsuperscript{1}, B. N. Harmon\textsuperscript{1}, R. J. McQueeney\textsuperscript{1} and A. I. Goldman\textsuperscript{1}}
\affiliation{\\\textsuperscript{1}Ames Laboratory, U.S. DOE and
Department of Physics and Astronomy\\
Iowa State University, Ames, IA 50011, USA}
\affiliation{\\\textsuperscript{2}Advanced Photon Source, Argonne
National Laboratory, Argonne, Illinois 60439, USA}

\date{\today}

\begin{abstract}
We describe x-ray resonant magnetic diffraction measurements at the
Fe $K$-edge of both the parent BaFe$_2$As$_2$ and superconducting
Ba(Fe$_{0.953}$Co$_{0.047}$)$_2$As$_2$ compounds.  From these
high-resolution measurements we conclude that the magnetic structure
is commensurate for both compositions. The energy spectrum of the
resonant scattering is in reasonable agreement with theoretical
calculations using the full-potential linear augmented plane wave
method with a local density functional.
\end{abstract}


\maketitle The observation of coexistence and competition between
superconductivity (SC) and antiferromagnetic (AFM) order in some
members of the iron arsenide family of superconductors has raised
interesting issues regarding the nature of both the SC and AFM
states. Several theoretical treatments have demonstrated that
coexistence is inconsistent with conventional BCS coupling, whereas
the s\textsuperscript{$\pm$} state, arising from pairing through
magnetic fluctuations, is compatible with coexistence and
competition between SC and AFM
order.\cite{Mazin_2008,Kuroki_2008,Chubukov_2008,Fernandes_2010,FandS_2010}
However, the nature of the AFM state in the doped superconducting
compounds, particularly the potential for incommensurabilty of the
magnetic structure, remains a significant issue under debate in both
theoretical and experimental work.

It has been argued that an incommensurate magnetic structure is
expected for the doped iron arsenides because of imperfect nesting
of the hole and electron Fermi surface
pockets,\cite{Vorontsov_2009,Cvetkovic_2009} referencing previous
work on chromium.\cite{Rice_1970} Some theoretical models find that
the coexistence between AFM and SC points to incommensurate AFM
order.\cite{Vorontsov_2009,Cvetkovic_2009,Lee_2010} However, it has
also been noted that while incommensurability may broaden the
coexistence regime,\cite{Vorontsov_2010} it does not appear to be a
prerequisite for coexistence.\cite{FandS_2010} Recent
calculations\cite{Park_2010} of the spin susceptibility in the
parent and doped $AE$Fe$_2$As$_2$ ($AE$ = Ca, Ba, Sr) compounds
point to incommensurability as the origin of the anisotropy observed
in the low-energy spin fluctuation spectrum of
Ba(Fe$_{0.926}$Co$_{0.074}$)$_2$As$_2$.\cite{HFLi_2010}

In contrast to the Fe$_{1+y}($Te$_{1-x}$Se$_x$)
family,\cite{LandD_2009,LandC_2010} all neutron diffraction
measurements to date indicate that the AFM order in the doped
$R$OFeAs ($R$ = rare earth) and $AE$Fe$_2$As$_2$ families is
commensurate,\cite{Cruz_08,Zhao_08_01,Huang_08,Chen_08,Zhao_08_02,Ryan_09,Lester_2009,Pratt_2009,Fernandes_2010,Kreyssig_2010}
and characterized by the so-called "stripe-like" magnetic
structure.\cite{LandD_2009,LandC_2010} Nevertheless, other
measurements employing local probes of magnetism, such as
\textsuperscript{75}As nuclear magnetic resonance
(NMR),\cite{Laplace_2009} muon spin relaxation
($\mu$SR),\cite{Carlo_2009} and \textsuperscript{57}Fe M\"{o}ssbauer
spectroscopy\cite{Bonville_2010} have proposed that the magnetic
order is, in fact, incommensurate for the doped compounds.
Zero-field $\mu$SR measurements on doped
LaFeAs(O$_{0.97}$F$_{0.03}$) noted a much faster damping of the
signal than found for the undoped parent compound and attributed
this to incommensurate AFM order.\cite{Carlo_2009} Supporting this
view, NMR measurements\cite{Ning_2009,Laplace_2009} on underdoped
Ba(Fe$_{1-x}$Co$_x$)$_2$As$_2$ ($x$ = 0.02, 0.04)\cite{Ning_2009}
and ($x$ = 0.06)\cite{Laplace_2009} found a strong broadening of the
\textsuperscript{75}As lines attributable to the appearance of a
distribution of internal fields at low temperatures in the
magnetically ordered state.  A quantitative comparison of the line
broadening for \textbf{H} $\parallel$ \textbf{c} and \textbf{H}
$\parallel$ \textbf{ab} led to the conclusion that there is a small
incommensurability, $\varepsilon$, in the magnetic structure such
that the commensurate propagation vector ($\frac{1}{2}$,
$\frac{1}{2}$, 1) in the undoped parent compound is given by
($\frac{1}{2} - \varepsilon$, $\frac{1}{2} - \varepsilon$, 1), with
$\varepsilon$ estimated to be approximately 0.04 reciprocal lattice
units (rlu), in the lightly Co-doped compounds.\cite{Laplace_2009}

To resolve this issue we present high-resolution x-ray resonant
magnetic scattering (XRMS) measurements at the Fe $K$-edge for two
samples; the parent BaFe$_2$As$_2$ compound and; Co-doped
Ba(Fe$_{0.953}$Co$_{0.047}$)$_2$As$_2$ which manifests coexistence
and competition between SC and AFM suggesting the possibility of
incommensurate magnetic order. We find that the magnetic Bragg peaks
are commensurate for both samples and scans along the [$\zeta$
$\zeta$ 0] and [$\zeta$ -$\zeta$ 0] directions allow us to place
limits on the magnitude of a potential incommensurability that are
much smaller than any value proposed to date. The energy spectrum of
the resonant scattering is in reasonable agreement with theoretical
calculations using the full-potential linear-augmented plane-wave
(FLAPW) method\cite{Blaha_2001} with a local density
functional.\cite{perdew_1992} These calculations suggest that the
resonant scattering at the Fe $K$-edge in the $\sigma$-to-$\pi$
scattering channel arises from dipole allowed transitions from the
core 1$s$ states to the unoccupied 4$p$ states that are spin
polarized due to hybridization with the 3$d$ states close to the
Fermi energy.

\begin{figure}
\begin{center}
\includegraphics[clip, width=.45\textwidth]{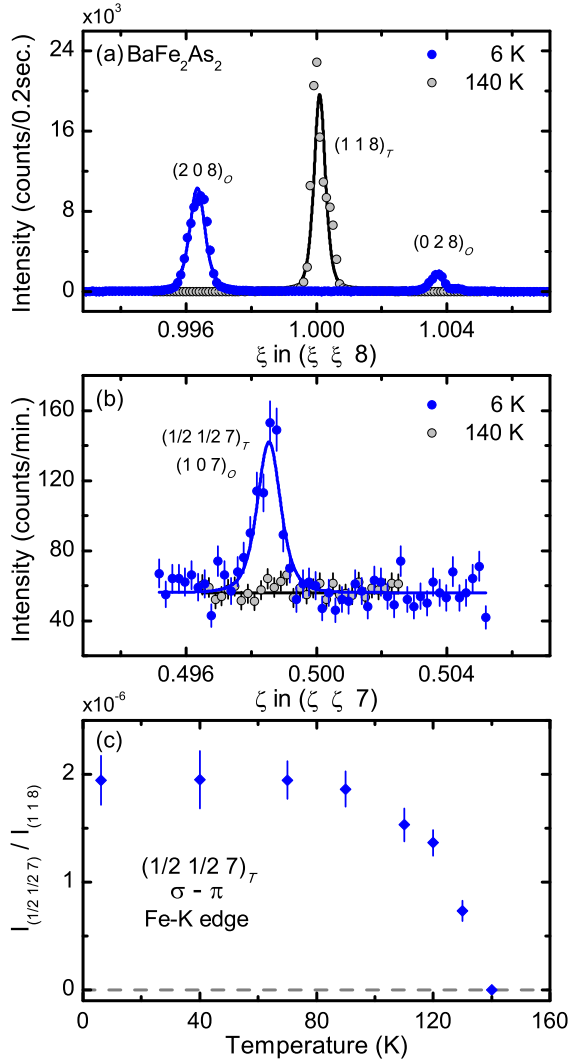}\\
\caption{(Color online) (a) Diffraction data from the parent
BaFe$_2$As$_2$ compound characteristic of the tetragonal structure
above, and the orthorhombic structure below, $T_S$ = 136 K. (b)
Scattering measured in the $\sigma$ - $\pi$ channel at the magnetic
Bragg peak position of the "stripe-like" AFM phase above and below
$T_N$ = $T_S$. (c) The temperature dependence of the integrated
intensity of the magnetic peak in (b) normalized to the (1 1 8)
charge reflection. } \label{parent}
\end{center}
\end{figure}

Single crystals of BaFe$_2$As$_2$ and
Ba(Fe$_{0.953}$Co$_{0.047}$)$_2$As$_2$ were grown out of a FeAs
self-flux using conventional high-temperature solution
growth.\cite{Ni_2008} Crystals from the same batch have been studied
by both neutron\cite{Pratt_2009,Fernandes_2010} and
x-ray\cite{Nandi_2010} scattering measurements previously. For the
XRMS measurements, pieces of the as-grown single crystals of
approximate dimensions 3 $\times$ 2 $\times$ 0.03
mm\textsuperscript{3} (BaFe$_2$As$_2$) and 7 $\times$ 4 $\times$
0.08 mm\textsuperscript{3} [Ba(Fe$_{0.953}$Co$_{0.047}$)$_2$As$_2$]
were selected. The extended surfaces of the crystals were
perpendicular to the \textbf{c} axis. The measured mosaicities of
the crystals were less than 0.02 degrees full-width-at-half-maximum
(FWHM), attesting to the high quality of the samples. The XRMS
experiment was performed on the 6ID-B beamline at the Advanced
Photon Source at the Fe $K$-edge ($E$ = 7.112 keV). The incident
radiation was linearly polarized perpendicular to the vertical
scattering plane ($\sigma$-polarized) with a spatial cross section
of 1.0 mm (horizontal) $\times$ 0.2 mm (vertical). In this
configuration, dipole resonant magnetic scattering rotates the plane
of linear polarization into the scattering plane
($\pi$-polarization). Cu (2 2 0) was used as a polarization analyzer
to suppress the charge and fluorescence background relative to the
magnetic scattering signal. For measurements of the magnetic
reflections, the sample was mounted at the end of the cold finger of
a displex cryogenic refrigerator with the tetragonal ($H H L$) plane
coincident with the scattering plane.

\begin{figure}
\begin{center}
\includegraphics[clip, width=0.45\textwidth]{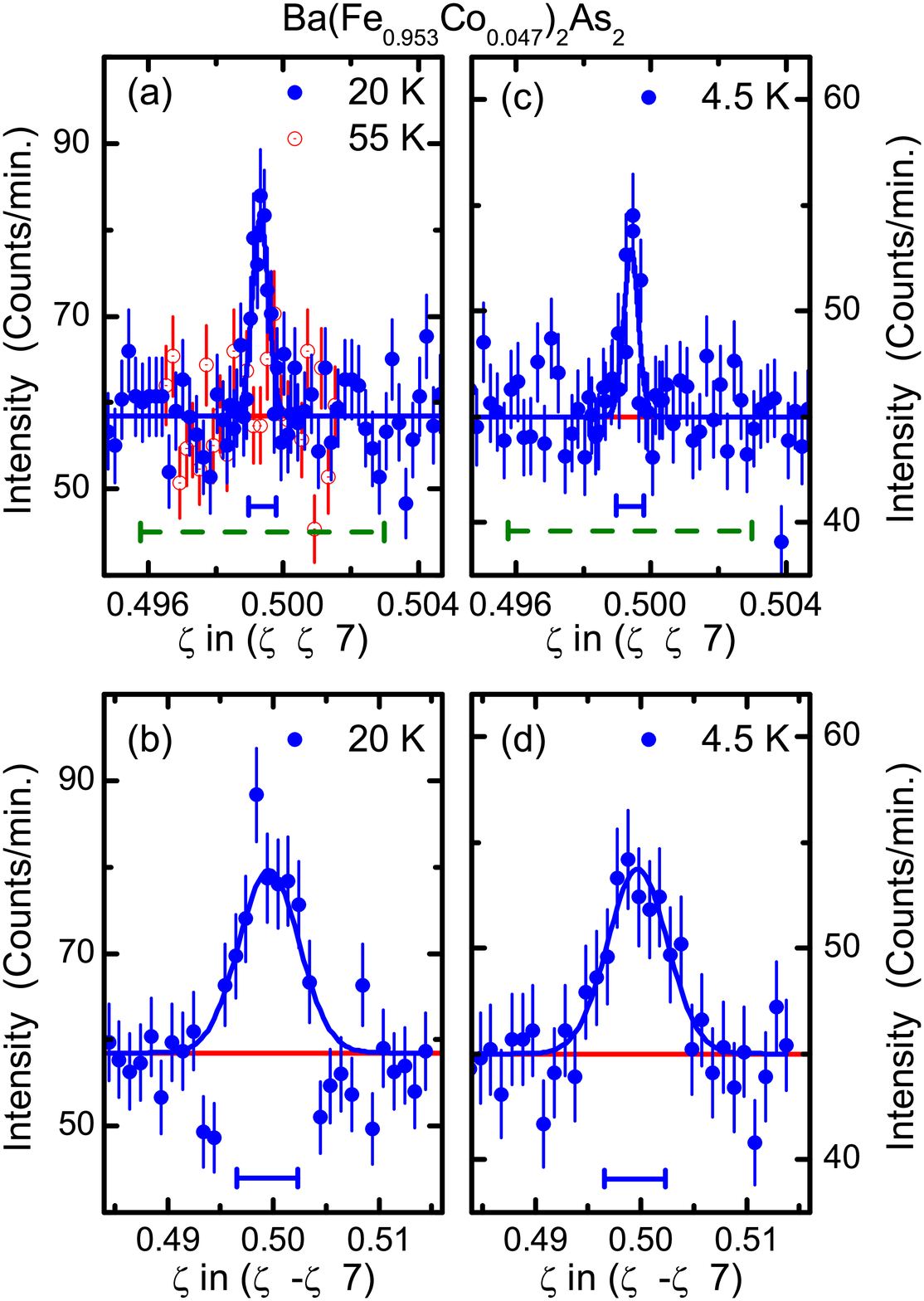}\\
\caption{(Color online) (a) [$\zeta$ $\zeta$ 0] scan through the
magnetic Bragg peak position of the "stripe-like" AFM phase at above
(55~K) and below (20~K) $T_N$ for the
Ba(Fe$_{0.953}$Co$_{0.047}$)$_2$As$_2$ sample. The solid bar
represents the experimental resolution for the x-ray measurements
along this direction while the dashed bar denotes the resolution of
our previous neutron measurements along this
direction.\cite{Pratt_2009} (b) [$\zeta$ -$\zeta$ 0] scan through
the magnetic Bragg peak position below $T_N$. The solid bar
represents the experimental resolution along this direction for our
x-ray measurements. The resolution width for neutron
measurements\cite{Pratt_2009} along [$\zeta$ -$\zeta$ 0] is a factor
of ten larger. (c) and (d) correspond to (a) and (b), respectively,
at the base temperature of 4.5~K.  The difference in the vertical
scale between panels (a),(b) and (c),d) arises from small
differences in the beam conditions for measurements performed
several months apart.} \label{cobalt}
\end{center}
\end{figure}

In Figs.~\ref{parent}(a) and (b) we display the raw data for the
parent BaFe$_2$As$_2$ compound for [$\zeta$ $\zeta$ 0] - scans
through the (1 1 8)$_{T}$ charge peak and ($\frac{1}{2}$
$\frac{1}{2}$ 7)$_{T}$ magnetic peak positions above and below the
coupled structural/magnetic transitions ($T_S$ = $T_N$ $\approx$ 136
K). These data were taken at the maximum in the resonant scattering
($E$ = 7.112 keV) at the Fe $K$-edge.  For temperatures below $T_S$
(=$T_N$), the charge peaks splits into the (2 0 8)$_O$ and (0 2
8)$_O$ peaks of the orthorhombic phase.  The disparity in
intensities is attributed to an imbalance in the domain populations
for these reflections within the illuminated volume of the sample.
Fig.~\ref{parent}(b) shows that, below $T_N$, scattering is clearly
observed at the (1 0 7)$_O$ magnetic peak position in the
orthorhombic phase with $a_O > b_O$. This diffraction peak arises
from magnetic domains characterized by the propagation vector (1 0
1)$_O$ or ($\frac{1}{2}$ $\frac{1}{2}$ 1)$_T$. Magnetic scattering
from domains characterized by the propagation vector (0 1 1)$_O$ or
($\frac{1}{2}$ $-\frac{1}{2}$ 1)$_T$ do not contribute to the
scattering in this geometry. For simplicity, we will henceforth
label all peaks with tetragonal indices. Therefore, (1 0 7)$_O$ will
be referred to as ($\frac{1}{2}$ $\frac{1}{2}$ 7)$_T$, keeping in
mind that the magnetic peaks are displaced from $\zeta$ =
$\frac{1}{2}$ because of the orthorhombic distortion. The measured
FWHM of 0.0007(1) rlu for the magnetic peak is the same (within
error) as that of the charge peak, consistent with long-range
magnetic order. Fig.~\ref{parent}(c) shows that as the sample
temperature increases, the intensity of the magnetic peak decreases
until it can no longer be observed above background at approximately
140 K, in agreement with previous neutron scattering
measurements.\cite{Huang_2008}

For the Ba(Fe$_{0.953}$Co$_{0.047}$)$_2$As$_2$ sample,
Fig.~\ref{cobalt} shows scans along the [$\zeta$ $\zeta$ 0] and
transverse [$\zeta$ -$\zeta$ 0] directions through the
($\frac{1}{2}$ $\frac{1}{2}$ 7)$_T$ magnetic Bragg peak position.
For the [$\zeta$ $\zeta$ 0] scan, the position of this peak is again
referenced to the (1 1 8)$_T$ charge peak and is displaced from
$\zeta$ = $\frac{1}{2}$ because of the orthorhombic distortion [see
Fig.~\ref{parent}(a)]. Along the [$\zeta$ $\zeta$ 0] direction
[Figs.~\ref{cobalt}(a) and (c)] and below $T_N$ = 47 K, we observe a
single peak, whereas an incommensurability of magnitude
$\varepsilon$ would result in two peaks split by 2$\varepsilon$. The
solid bar beneath the data in Figs.~\ref{cobalt}(a) and (c)
describes the measured FWHM of the (1 1 8)$_T$ charge peak and
represents our experimental resolution along [$\zeta$ $\zeta$ 0].
Therefore, the FWHM of 0.0007(1) rlu for the ($\frac{1}{2}$
$\frac{1}{2}$ 7)$_T$ magnetic Bragg peak along this direction places
an upper limit on the potential incommensurability ($\varepsilon
\thickapprox 3.5~X~10^{-4}$) which is two orders of magnitude
smaller than the value proposed in Ref. [\onlinecite{Laplace_2009}].
We have also checked along the transverse [$\zeta$ -$\zeta$ 0]
direction for any evidence of incommensurability as shown in
Figs.~\ref{cobalt}(b) and (d). However, for the present experimental
configuration, our resolution along this direction is coarser
[0.0067(15) rlu]. Nevertheless, these data still allow us to place
an upper limit on the incommensurability ($\varepsilon \thickapprox
3.3~X~10^{-3}$) that is more than an order of magnitude smaller than
that proposed.\cite{Laplace_2009} Furthermore, a comparison of the
scans at 20~K and 4.5~K show that there is no evidence of additional
line broadening for this compound below the superconducting
transition ($T_c$ = 17~K).

The dashed bars in Figs.~\ref{cobalt}2(a) and (b) represent the
experimental resolution for our previous neutron diffraction
measurements on Ba(Fe$_{0.953}$Co$_{0.047}$)$_2$As$_2$ along the
[$\zeta$ $\zeta$ 0] direction.\cite{Pratt_2009} Even with the poorer
resolution of this measurement, an incommensurability of
$\varepsilon$ = 0.04 rlu would have been clearly observed in scans
performed along the [$\zeta$ $\zeta$ 0] direction. Our XRMS
measurements, however, now place a strong limit on the magnitude of
any incommensurability for the Co-doped compound. In this light, the
broadened lineshapes measured by $\mu$SR,\cite{Carlo_2009}
NMR,\cite{Laplace_2009} and M\"{o}ssbauer
spectroscopy\cite{Bonville_2010} must arise from other causes.
Density-functional theory calculations by Kemper \emph{et
al.}\cite{Kemper_2010} indicate that although the nonmagnetic
scattering potential associated with Co-doping in the FeAs planes is
relatively well localized, the magnetic potential significantly
perturbs the spin density wave state over much longer length scales.
This, in turn, leads to a large distribution of hyperfine fields, as
pointed out by Dioguardi \emph{et al.},\cite{Dioguardi_2010} who
suggest that the origin of the broadening in their NMR studies of
Co- and Ni-doped BaFe$_2$As$_2$ is consistent with doping-induced
disorder in the AFM state rather than incommensurate order.

\begin{figure}
\begin{center}
\includegraphics[clip, width=0.45\textwidth]{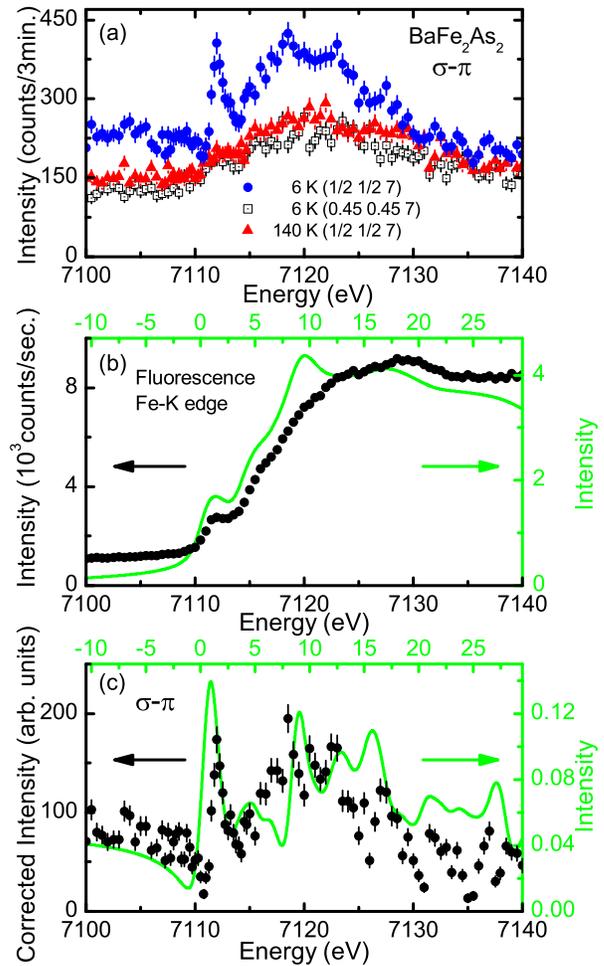}\\
\caption{(Color online) (a) Energy scans through the ($\frac{1}{2}$
$\frac{1}{2}$ 7)$_T$ magnetic peak above (filled triangles) and
below (filled circles) $T_N$, and at low temperature away from the
magnetic Bragg peak (open squares) (b) The measured fluorescence
(filled circles) and calculated absorption (line) as described in
the text. (c) The background subtracted and absorption corrected
XRMS signal (filled circles) along with the calculated XRMS spectrum
(line).} \label{resonant}
\end{center}
\end{figure}

We now turn to a description of the energy spectrum associated with
the resonant scattering from BaFe$_2$As$_2$.  In
Fig.~\ref{resonant}(a) we show the raw data from energy scans at
constant \textbf{Q} = ($\frac{1}{2}$ $\frac{1}{2}$ 7)$_T$ at $T$ = 6
K (filled circles), well below $T_N$.  To determine the background
at this scattering vector, energy scans were also performed at
($\frac{1}{2}$ $\frac{1}{2}$ 7)$_T$ for $T$ = 140 K (red triangles),
just above $T_N$, and at \textbf{Q} = (0.45 0.45 7), away from the
magnetic peak, at $T$ = 6 K (open squares). The shape of the
background in the vicinity of the Fe $K$-edge is consistent with an
increase in the fluorescence from the sample
[Fig.~\ref{resonant}(b)]. The background subtracted and absorption
corrected energy scan at ($\frac{1}{2}$ $\frac{1}{2}$ 7)$_T$ shown
in Fig.~\ref{resonant}(c) contains several components: (1) an energy
independent contribution that is most clearly visible below the
absorption edge; (2) a noticeable dip in the scattering intensity
followed by; (3) a sharp feature close to the absorption threshold
and broad scattering that extends to energies more than 20 eV above
the absorption edge. This energy spectrum is similar to the one
observed in previous XRMS measurements in the $\sigma$-$\pi$
scattering channel at the Ni $K$-edge for NiO.\cite{Neubeck_2001}
The energy independent scattering contribution (1) arises from
nonresonant magnetic scattering while the resonant features (3) at
and above the Fe $K$-edge can be attributed to dipole ($E1$)
transitions from the 1$s$ initial state to the unoccupied 4$p$
states that are weakly polarized through hybridization with 3$d$
states near the Fermi energy.  The sharp feature close to the
absorption threshold may also contain a contribution from quadrupole
($E2$) allowed transitions from the 1$s$ to 3$d$ states, but a clear
separation of the $E1$ and $E2$ contributions will require further
measurements of the angular dependence of the scattering as well as
the $\sigma$-$\sigma$ scattering channel. The dip in the scattering
(2) arises from interference between the nonresonant and resonant
magnetic scattering as the phase of the resonant scattering changes
across the absorption edge.

To model the resonant scattering energy scans, we have used a
full-potential linear augmented plane wave (FLAPW)
method\cite{Blaha_2001} with a local density
functional.\cite{perdew_1992} Details of the calculations will be
presented elsewhere, and only briefly outlined here. To obtain a
self-consistent charge and potential, we chose 810 $\textbf{k}$
points in the irreducible Brillouin zone (IBZ), and set
$\textbf{R}_{MT}$*$\textbf{k}_{max}$ = 8.0, where R$_{MT}$ is the
smallest muffin-tin radius and $\textbf{k}_{max}$ is the basis set
cutoff (the maximum value of $|\textbf{k}+\textbf{K}_i|$ included in
the basis). The muffin-tin radii are 2.4, 2.2, 2.2 a.u. for Ba, Fe,
and As respectively. The self-consistent calculation was iterated
until the total energy convergence reached 0.01 mRy/cell. For the
x-ray absorption spectra [Fig.~\ref{resonant}(b)] and XMRS
[Fig.~\ref{resonant}(c)] we calculated empty states up to 40 eV
above Fermi energy with 1320 $\textbf{k}$ points in IBZ and with the
calculated self-consistent potential. Our calculation of the $E2$
contribution to the sharp feature close to the absorption threshold
indicates that it is much smaller than the $E1$ contribution.  To
model the interference between the resonant and nonresonant
scattering close to the absorption edge, an energy-independent
scattering amplitude, equal to the resonant scattering contribution
was added to the real part of the resonant scattering amplitude. The
calculated energy spectrum was broadened with a 1.25 eV
Lorentzian\cite{Krause_1979} to account for the core-hole life time,
a 1 eV Gaussian for the instrumental resolution. The calculated
absorption and resonant scattering spectra are displayed as lines in
Fig.~\ref{resonant}(b) and Fig.~\ref{resonant}(c), respectively, and
capture the essential features of our measurements including
features (1)-(3) discussed above.

In summary, we have used XRMS at the Fe $K$-edge to directly probe
the commensurability of the magnetic structure in
Ba(Fe$_{0.953}$Co$_{0.047}$)$_2$As$_2$ with high resolution. The AFM
structure is commensurate and the FWHM of scans measured along the
[$\zeta$ $\zeta$ 0] direction places an upper limit on the potential
incommensurability which is two orders of magnitude smaller than the
value proposed in Ref. [\onlinecite{Laplace_2009}]. Energy scans
through the resonance at the Fe $K$-edge are in reasonable agreement
with theoretical calculations and these calculations suggest that
the resonant scattering at the Fe $K$-edge in the $\sigma$-to-$\pi$
scattering channel arises from dipole allowed transitions from the
core 1$s$ states to the unoccupied 4$p$ states that are spin
polarized due to hybridization with the 3$d$ states close to the
Fermi energy.

We acknowledge valuable discussions with J. Lang, J. Schmalian and
R. M. Fernandes.  The work at Ames Laboratory was supported by the
US DOE Office of Science, Basic Energy Sciences under Contract No.
DE-AC02-07CH11358. Use of the Advanced Photon Source was supported
by the US DOE under Contract No. DE-AC02-06CH11357.

\bibliographystyle{apsrev}
\bibliography{resonant_ba122_arxiv}

\end{document}